# Electron dynamics in laser and quasi-static transverse electric and longitudinal magnetic fields

Yanzeng Zhang and S. I. Krasheninnikov
University of California San Diego, 9500 Gilman Dr., La Jolla, CA 92095

**Abstract.**

By deriving the 3/2 dimensional Hamiltonian equations for electrons in the intense laser radiation and quasi-static transverse electric and longitudinal magnetic fields, the electron heating mechanisms are examined both for low harmonic resonance of electron frequency in the static fields with the laser frequency and for high harmonic resonances where the overlapping of broadened resonances causes the stochastic heating. For both cases, the maximum electron kinetic energies, well beyond the ponderomotive scaling, depend only on a small parameter combining the laser amplitude, electrostatic field strength and the conserved dephasing rate.

## 1. Introduction

The generation of energetic electrons through direct laser acceleration with the assistance of self-generated or externally applied quasi-static electric and magnetic fields remains as an active research field in the course of laser plasma interactions (e.g., see [1-5]). In this process, the resonance between electron motion in the transverse electric [6, 7] or magnetic [8] field with laser pulse can significantly accelerate electrons. The reduction of electron dephasing rate due to the presence accelerating [9, 10] and the stochastic motion [7, 11] in the decelerating longitudinal (along laser propagating direction) electric field can also facilitate the electron acceleration. On the other hand, the generation of strong spontaneous longitudinal magnetic field in the laser plasma interaction [12-14] has made the resonant acceleration of electrons, which requires a matching condition for electron gyrofrequency and laser frequency, possible [15, 16]. All these mechanism can accelerate the electron with energy well beyond the ponderomotive scaling.

However, due to the multidimensionality of the system and the strong nonlinearity of the relativistic electron dynamics, the fully theoretical analysis of electron dynamics in the synergistic quasi-static and laser fields is still an interesting problem. Recently, it was shown that the electron dynamics in an arbitrarily polarized laser radiation and transverse or longitudinal quasi-static electric fields as well as transverse magnetic field can be described by the 3/2 dimensional (3/2D) Hamiltonian approach [17, 18], which greatly simplifies the analysis of electron interactions with laser and quasi-static electromagnetic fields. It was demonstrated [19] that, the acceleration of electron is attributed to an onset of stochasticity [20, 21].

In the present paper, we devote to extending the 3/2D Hamiltonian approach to the case of electron interactions with laser pulse and quasi-static longitudinal magnetic and transverse electric fields (here we use attractive electrostatic potential U). Two cases will be considered: one is for weak electric but strong magnetic fields such that the electron gyrofrequency in the magnetic field and laser frequency are approximately matched and electron is resonantly accelerated. The second case is stochastic acceleration of the electron in both weak electric and magnetic fields, where high-harmonics resonances of unperturbed electron oscillation frequency in the quasi-static fields and laser frequency can overlap. For convenience, in the following, we will use the dimensionless variables, where t and $\vec{r}$ are normalized, respectively by the laser wave frequency ($\omega$) and $\omega/c$, the vector potential of laser and quasi-static magnetic field by $e/m_e c^2$ and the electrostatic potential by $e/m_e c$, where e is the elementary charge, $m_e$ is the electron mass and c is the speed of light in vacuum

The remainder of the paper is organized as follows. In section 2 we will derive the 3/2D Hamiltonian of the electron in the laser pulse and transverse electric and longitudinal magnetic fields and discuss the electron trajectories in these static fields with an impact of the laser radiation. The efficient electron heating via single low harmonic resonance is examined in section 3, while section 4 investigates the electron dynamics for weak static fields, where overlapping of the broadened high harmonic resonances of electron frequency with the laser frequency would result in stochastic electron heating. In section 6 the main results are summarized

## 2. 3/2D Hamiltonian and electron trajectories

In this section, we derive the 3/2D Hamiltonian for electron in the electromagnetic field described by the vector potential in the form of

$$\vec{A} = \tilde{\vec{A}} - \vec{e}_y t \partial U(y)/\partial y + \vec{e}_x A_B(y), \tag{1}$$

where the electrostatic potential, $U(y)$, and magnetic vector potential, $A_B(y)$, depend only on the coordinate y illustrating, respectively, the transverse electric and longitudinal magnetic fields; whereas an arbitrarily polarized laser wave is denoted by the vector potential $\tilde{\vec{A}} = \tilde{A}_x(\xi)\vec{e}_x + \tilde{A}_y(\xi)\vec{e}_y$ with $\tilde{A}_x(\xi)$ and $\tilde{A}_y(\xi)$ being arbitrary functions of the variable $\xi \equiv t - z/\alpha$ ( $\alpha = v_{ph}/c \geq 1$ is the normalized laser phase velocity). Note that $\vec{E}_{stat.} = \nabla U$ by using Eq. (1) in order to denote an attractive potential. Then, it is easy to show that there are two integrals of electron motion

$$\bar{P}_x \equiv P_x - (\tilde{A}_x + A_B), \tag{2}$$

$$C_\perp \equiv \gamma - \alpha P_z + U(y), \tag{3}$$

where $P_{x,z}$ are the (x, z)-component of electric momentum and $\gamma \equiv \sqrt{1+P^2}$ is the relativistic factor (normalized electron kinetic energy).

Introducing the variable $\tilde{p}_y = P_y - \tilde{A}_y$, after some algebra similar to that in [18], the 3/2D Hamiltonian equations read

$$\frac{d\tilde{p}_y}{d\xi} = -\frac{\partial H}{\partial y}, \text{ and } \frac{dy}{d\xi} = \frac{\partial H}{\partial \tilde{p}_y}, \tag{4}$$

where

$$H(\tilde{p}_y, y, \xi) = \frac{\alpha}{\alpha^2-1}\left\{\sqrt{(U-C_\perp)^2 + (\alpha^2-1)P_{\perp,y}^2} + \alpha U - \frac{C_\perp}{\alpha}\right\} \equiv \gamma + U = E, \tag{5}$$

and

$$P_{\perp,y}^2 = 1 + (\bar{P}_x + \tilde{A}_x + A_B)^2 + (\tilde{p}_y + \tilde{A}_y)^2. \tag{6}$$

Notice that the Hamiltonian in equation (5) corresponding to the luminal phase velocity $\alpha \to 1$ becomes

$$H = \frac{1}{2}\left\{\frac{1+(\bar{P}_x + \tilde{A}_x + A_B)^2 + (\tilde{p}_y + \tilde{A}_y)^2}{C_\perp - U(y)} + U(y) + C_\perp\right\} \equiv \gamma + U(y) = E = P_z + C_\perp. \tag{7}$$

The strongest interaction between the electron and laser wave occurs for the case where electron stays in phase with the laser wave [19], which implies small difference $\gamma - P_z$. However, from Eq. (3) we have $\gamma - P_z = C_\perp - U(y) + (\alpha - 1)P_z$, which indicates that whereas static electric field could enhance the electron-laser interaction when U approaches $C_\perp$, the superluminal phase velocity of laser radiation, $\alpha > 1$, would reduce it [19].

For simplicity, in the rest of this paper, we consider the luminal case, $\alpha = 1$, and assume that $\bar{P}_x = 0$. To specify static transverse electric and longitudinal magnetic fields we take $U = \kappa y^2/2$ and $A_B = B_0 y$ where $\kappa$ and $B_0$ are some constants (we take $\kappa > 0$ such that the electric field will force the electron towards the axis). This is equivalent to the setup of constant longitudinal magnetic field and electric field with linear dependence on the transverse coordinate. This choice of electric field is widely used and the dimensionless parameter $\kappa$ for self-generated electric field is dependent on the ion density in the ion-channel, $\kappa = \omega_{pe}^2/\omega^2$, (e.g., see [3]), where $\omega_{pe} = \sqrt{4\pi n_0 e^2/m_e}$ is the plasma frequency. Taking into account the dephasing rate is positive, $\gamma - P_z = C_\perp - U(y) > 0$, such choice of electrostatic potential guarantees that $C_\perp > 0$. Moreover, we will use linearly polarized planar laser wave, i.e., either $\tilde{\vec{A}} = a_0 \sin(\xi)\vec{e}_y$ or $\tilde{\vec{A}} = a_0 \sin(\xi)\vec{e}_x$ where $a_0$ is the normalized amplitude of laser vector potential.

We first consider electron trajectory neglecting an impact of the laser field, where the Hamiltonian (7) is conserved, $H = \text{const.}$, and from Eqs. (4, 7) we find

$$y = \sqrt{2EC_\perp(\kappa E + B_0^2)^{-1}} \cos\theta, \quad \tilde{p}_y = -\sqrt{2EC_\perp}\sin\theta, \tag{8}$$

where angle $\theta$ (e.g., see Fig. 1) is determined by

$$\frac{d\theta}{d\xi} = \frac{(\kappa E + B_0^2)^{3/2}}{(B_0^2 + \kappa E \sin^2\theta)C_\perp}. \tag{9}$$

In Eq. (8), we choose the origin of $\theta$ such that $\theta = 0$ and $\theta = \pi$ correspond to, respectively, the maximum and minimum of the coordinate y: $y = y_{max}$, $y = y_{min}$ as shown in Fig. 1. Integrating Eq. (9) one can find

$$\xi = \frac{C_\perp}{(\kappa E + B_0^2)^{3/2}}\left[B_0^2\theta + \frac{2\theta - \sin(2\theta)}{4}\kappa E\right] + \text{const.}. \tag{10}$$

It follows that the period of unperturbed electron motion is $T = \xi(\theta = 2\pi) - \xi(\theta = 0)$ and thus the oscillating frequency, $\Omega = 2\pi/T$, reads

$$\Omega = \frac{(\kappa E + B_0^2)^{3/2}}{C_\perp}\left(B_0^2 + \frac{\kappa E}{2}\right)^{-1}, \tag{11}$$

which can also be obtained from $\Omega = \partial E/\partial I$, where $I = \oint \tilde{p}_y dy/2\pi = EC_\perp(\kappa E + B_0^2)^{-1/2}$ is the action of electron oscillation. However, an impact of laser wave on electron trajectory could be largely ignored, from Eq. (7) only for the energies [19]

$$E > E_{pond} = a_0^2/2C_\perp, \tag{12}$$

where $E_{pond}$ is considered as the ponderomotive energy scaling.

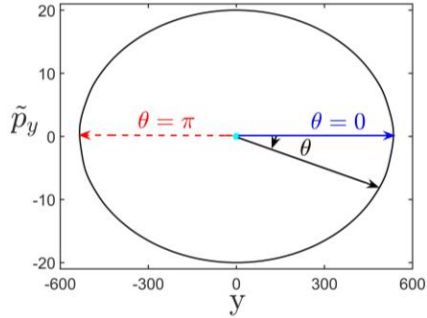

Fig. 1. Schematic view of the ultrarelativistic electron trajectory on ($\tilde{p}_y$, y) plane for $C_\perp = 1$, $\kappa = 10^{-5}$, $B_0 = 0.02$ and $E = 100$.

Recalling that according to our normalization convention, frequency $\Omega$ is normalized to the laser frequency, we conclude that there are two candidate mechanisms for efficient electron acceleration: One is the low harmonics resonance of electron frequency $\Omega$ with laser frequency [15, 16], $n\Omega = 1$ (where $n \gtrsim 1$ is the integer number); while the second is resonance broadening results in the overlap of the resonances for $n \gg 1$, which causes stochastic electron heating [19]. However, no matter which mechanism is active, the requirement $\Omega \leq 1$ will limit the electron acceleration considering that at $\Omega > 1$ an impact of laser field becomes adiabatic. In the following two sections, we will examine the electron dynamics in these two regimes

From Eq. (11) one can see that the inequality $\Omega > 1$ requires $B_0$, E or both of them are large. Especially, for

$$B_0 > C_\perp \quad (13)$$

$\Omega > 1$ is true for all E, whereas for

$$E > E_{max}^{abs} = C_\perp^2 / 4\kappa, \quad (14)$$

$\Omega > 1$ is satisfied for all $B_0$. Therefore, $E_{max}^{abs}$ could be considered as an absolute maximum energy that an electron can be gained via interaction with laser wave, and a small parameter

$$\varepsilon \equiv \sqrt{E_{pond} / E_{max}^{abs}} = \sqrt{2} a_0 \kappa^{1/2} C_\perp^{-3/2} < 1, \quad (15)$$

should be considered for electron being accelerated beyond the ponderomotive scaling.

From Eqs. (4, 7), the electron energy variations, $E(\xi) = \int^\xi (\partial H / \partial \xi) d\xi$, due to an impact of the laser radiation are

$$E_y(\xi) - E_y(\xi_{max}) = \int_{\xi_{max}}^{\xi} \frac{a_0 (a_0 \sin\xi + \tilde{p}_y) \cos\xi}{C_\perp - U(y)} d\xi, \quad (16a)$$

and

$$E_x(\xi) - E_x(\xi_{max}) = \int_{\xi_{max}}^{\xi} \frac{a_0 [a_0 \sin\xi + A_B(y)] \cos\xi}{C_\perp - U(y)} d\xi, \quad (16b)$$

for $\vec{\tilde{A}} = a_0 \sin(\xi) \vec{e}_y$ and $\vec{\tilde{A}} = a_0 \sin(\xi) \vec{e}_x$, respectively.

## 3. Low harmonic resonance between electron and laser frequencies

In this section, we will examine the electron dynamics for low-n resonance. If $\kappa = 0$, the resonant (matching) condition is given by $B_0 = C_\perp$, which, independent with electron energy, usually requires a strong longitudinal magnetic field for $C_\perp \sim 1$ (e.g., for laser wavelength of $\lambda = 1\mu m$, the matching condition is satisfied for $B_0 \sim 10 kT$). Fortunately, the pre-acceleration of electron in the direction of laser propagating ($C_\perp < 1$) could relax this limitation [16]. However, in the presence of quasi-static electric field, the acceleration of electron will lead to the departure of electron from the matching condition as seen from Eq. (11). If electron is accelerated to case where $B_0^2 \ll \kappa E$, the further acceleration will be terminated. Therefore, to maximize the efficiency of electron heating, it is important that the frequency $\Omega = 1/n$ is dominated by the magnetic field and we will consider the case of $B_0^2 \gg \kappa E$ even for the maximum electron energy. As a result, we have $\Omega \approx B_0/C_\perp$, and thus the initial matching condition remains still great importance.

Consider the condition of $\Omega \sim 1$, the electron laser interaction could be effective along the whole orbit shown in Fig. 1 for $E > E_{pond}$. Using the expansion of

$$e^{iz\sin\theta} = \sum_{m=-\infty}^{\infty} J_m(z)e^{im\theta}, \tag{17}$$

where $J_{-m}(z) = (-1)^m J_m(z)$ is the $m$-th Bessel function of the first kind [22], we obtain the energy variation in Eq. (16) as following:

$$\Delta E_y \approx \frac{\pi a_0 (EC_\perp)^{1/2}}{2^{1/2} B_0} \sum_m J_{m=(\Omega_B^{-1}\pm 1)/2}(\rho)\sin\xi_j + \frac{\pi a_0^2}{2B_0}\sum_m J_{m=\Omega_B^{-1}}(2\rho)\sin(2\xi_j), \tag{18a}$$

$$\Delta E_x \approx \frac{\pi a_0 (EC_\perp)^{1/2}}{2^{1/2} B_0} \sum_m J_{m=(\Omega_B^{-1}\pm 1)/2}(\rho)\cos\xi_j + \frac{\pi a_0^2}{4B_0}\sum_m J_{m=\Omega_B^{-1}}(2\rho)\sin(2\xi_j). \tag{18b}$$

where

$$\rho = (4\Omega_E)^{-1} \ll 1, \quad \Omega_E = \frac{(\kappa E + B_0^2)^{3/2}}{C_\perp \kappa E} \gg 1, \text{ and } \Omega_B = \frac{(\kappa E + B_0^2)^{3/2}}{C_\perp B_0^2} \sim 1. \tag{19}$$

Considering the property of Bessel function as $\rho \to 0$, the efficient electron energy gain is only possible for m=0 ($\Omega_B = 1$ and thus $C_\perp = B_0$), where the difference of the energy change due to the laser polarization has disappeared.

From equations (18) we see that $\Delta E \propto E^{1/2}$ and thus it will continuously increase to infinity if the transverse electric field disappears. However, the presence of the transverse electric field, which induces a correction to the electron frequency, $|\Delta\Omega| \approx \kappa E/B_0^2$, will set an upper limit to the resonant electron energy since after $m \simeq 1/|\Delta\Omega|$ circles of cyclotron motion the sign of the energy gain (due to $\sin\xi_j$ or $\cos\xi_j$) in equation (18) changes. Then the maximum of the electron energy for $\Omega_B = 1$ can be estimated as

$$E_{max}^{\Omega_B=1} \approx \frac{\pi^2 a_0^2 C_\perp}{2^3 B_0^2} \times m^2. \tag{20}$$

Noticing that $\Omega$ depends on the electron energy, we can approximate m by using the averaged electron energy, $\bar{E} \approx E_{max}^{\Omega_B=1}/2$, such that $m \approx B_0^2/2\kappa\bar{E}$. As a result, the maximum electron energy obtained from equation (20) reads

$$E_{max}^{\Omega_B=1} \approx (2\pi)^{2/3} E_{max}^{abs} \varepsilon^{2/3} = (2\pi)^{2/3} E_{pond} \varepsilon^{-4/3}. \tag{21}$$

It follows that, for the small parameter $\varepsilon \ll 1$, this resonant energy is smaller than $E_{max}^{abs}$ but larger than the ponderomotive scaling.

To confirm the energy scaling in Eq. (21), we have numerically integrated the Hamiltonian equations in Eqs. (4, 7) and the results are shown in Fig. 2. In Fig. 2(a), we have employed different parameters of $B_0 = C_\perp$, $a_0$ and $\kappa$ for the initial condition of $y = 0$ and $\tilde{p}_y = 0$, which is in great agreement with Eq. (21). Fig. 2(b) shows the impact of initial conditions on the evolution of electron energy, from which we see that the initial conditions have almost no effect on the energy gain but only shift the profile. The typical electron orbit in the phase space is sketched in Fig. 2(c).

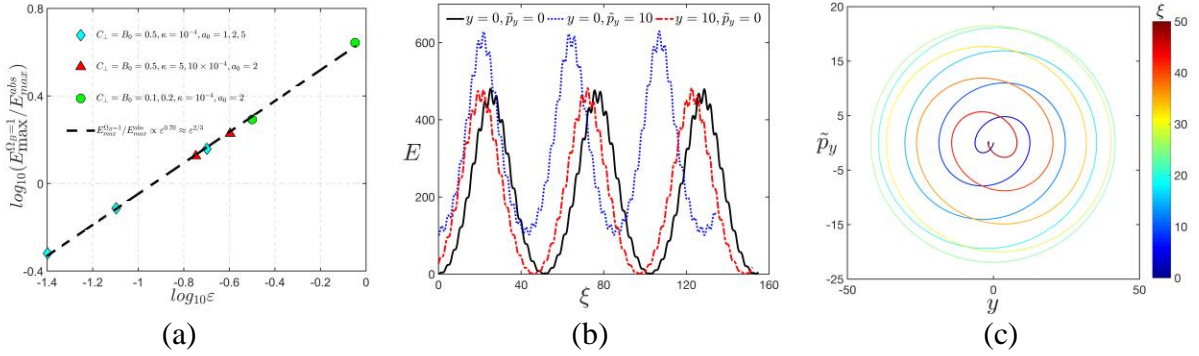

Figure 2. (a) Maximum electron energy scaling for Eq. (21) with initial conditions $y = 0$, $\tilde{p}_y = 0$ and different parameters of $B_0 = C_\perp$, $a_0$, and $\kappa$; (b) Evolution of E for $B_0 = C_\perp = 0.5$, $a_0 = 2$, $\kappa = 10^{-4}$ and different initial conditions ($y, \tilde{p}_y$); (c) Schematic view of electron trajectories for $B_0 = C_\perp = 0.5$, $a_0 = 2$, $\kappa = 10^{-4}$ and initial conditions $y = 0$, $\tilde{p}_y = 0$, where color bar shows the evolution of time $\xi$.

## 4. High-n resonances and stochastic electron heating

In this section, we examine the case where $\Omega \ll 1$ (thus $\kappa E + B_0^2 \ll C_\perp^2$) and thus electron heating is possible via overlapping of high-n resonances (stochastic heating). One main characteristic of stochastic motion is that the Hamiltonian (electron energy) variation occurs only in a relatively short time compared with the electron oscillation period and from Eq. (7) this nonadiabatic interaction ("kick") takes place in the vicinity of $y = y_{max}$ and $y = y_{min}$. If we assume that electron energy variation, $\Delta E$, during single kick is small $|\Delta E| \ll E$ [8], the unperturbed electron trajectories in equations (8, 9) can be applied to assess the electron energy change between two consecutive collisions in Eq. (16), yielding

$$\Delta E_y = a_0 \left( \frac{2EC_\perp}{\kappa E + B_0^2} \right)^{1/2} \sin \xi_j \int_{-\pi/2}^{\pi/2} \sin\theta \sin\left[ \frac{\theta}{\Omega_B} + \frac{2\theta - \sin(2\theta)}{4\Omega_E} \right] d\theta \\ + \frac{a_0^2}{2(\kappa E + B_0^2)^{1/2}} \sin(2\xi_j) \int_{-\pi/2}^{\pi/2} \cos\left[ \frac{2\theta}{\Omega_B} + \frac{2\theta - \sin(2\theta)}{2\Omega_E} \right] d\theta$$  (22a)

$$\Delta E_x = a_0 B_0 \frac{(2EC_\perp)^{1/2}}{\kappa E + B_0^2} \cos \xi_j \int_{-\pi/2}^{\pi/2} \cos\theta \cos\left[ \frac{\theta}{\Omega_B} + \frac{2\theta - \sin(2\theta)}{4\Omega_E} \right] d\theta \\ + \frac{a_0^2}{2(\kappa E + B_0^2)^{1/2}} \sin(2\xi_j) \int_{-\pi/2}^{\pi/2} \cos\left[ \frac{2\theta}{\Omega_B} + \frac{2\theta - \sin(2\theta)}{2\Omega_E} \right] d\theta$$  (22b)

where $\xi_j$ is the "time" of previous "collision".

In [19] it was shown that the transverse electric filed itself leads to stochastic electron heating, depending on laser polarization, up to the energies

$$E_{max}^y (B_0 = 0) \sim E_{max}^{abs} \varepsilon^{6/7} \quad \text{and} \quad E_{max}^x (B_0 = 0) \sim E_{max}^{abs} \varepsilon^{12/11}.$$  (23)

We note that for $\varepsilon \ll 1$ both of these energies are below $E_{max}^{abs}$ and even smaller than $E_{max}^{\Omega_B=1}$ in Eq. (21), but above the ponderomotive scaling. However, the presence of $B_0$ will change the electron dynamics.

Taking into account that $|\theta|\ll 1$ mostly contributes to the integrals in equation (23) under the condition of $\Omega\ll 1$, we can use Taylor expansion of the terms in the square brackets of equations (22)

$$\Omega_B^{-1}\theta+\Omega_E^{-1}\left[2\theta-\sin(2\theta)\right]/4 \approx \Omega_B^{-1}\theta+\Omega_E^{-1}\theta^3/3, \qquad (24)$$

and the integral limit can be extended to infinity. As a result, the integrals in equations (16a) and (16b) are degenerated to the Airy function $Ai(x)$ and its first derivative $Ai'(x)$, i.e.,

$$\Delta E_y \approx -2\pi a_0 \left(\frac{2EC_\perp}{\kappa E+B_0^2}\right)^{1/2}\Omega_E^{2/3}\sin\xi_j Ai'(\eta)+\frac{\pi a_0^2}{(\kappa E+B_0^2)^{1/2}}2^{-1/3}\Omega_E^{1/3}\sin(2\xi_j)Ai(2^{2/3}\eta), \qquad (25a)$$

$$\Delta E_x = 2\pi a_0 B_0 \frac{(2EC_\perp)^{1/2}}{\kappa E+B_0^2}\Omega_E^{1/3}\cos\xi_j Ai(\eta)+\frac{\pi a_0^2}{(\kappa E+B_0^2)^{1/2}}2^{-1/3}\Omega_E^{1/3}\sin(2\xi_j)Ai(2^{2/3}\eta), \qquad (25b)$$

where $\eta\equiv\Omega_E^{1/3}\Omega_B^{-1}=B_0^2 C_\perp^{2/3}(\kappa E)^{-4/3}/(1+B_0^2/\kappa E)$. We note that the results in Eq. (25) recover the results presented in [19] for $B_0=0$. Considering the exponential decaying property of $Ai(x)$ and $Ai'(x)$ for $x>1$, we see that efficient electron energy gain occurs for $\eta\lesssim 1$. Note that efficient stochastic heating requires the electron frequency strongly depending on electron energy and thus we consider the case of $B_0^2\ll \kappa E$ such that $\Omega\approx\Omega_E$.

Ignoring the differences induced by $Ai(\eta)$, $Ai'(\eta)$ and $Ai(2^{2/3}\eta)$ which are in the same order, the condition of $|\Delta E|\ll E$ and thus $E\gg E_{max}^{abs}\varepsilon^{3/2}=E_{pond}\varepsilon^{-1/2}$ as seen from Eq. (25) guarantees that the first part on the right hand side (RHS) of equation (25a) dominates and as a result, we obtain

$$\Delta E_y \approx -2^{7/3}\pi Ai'(\eta)(E_{max}^{abs})^{2/3}\varepsilon E^{1/3}\sin(\xi_j), \qquad (26)$$

and $\eta\approx (B_0/C_\perp)^2(4E_{max}^{abs}/E)^{4/3}$.

The time interval between two consecutive kicks can be approximated by half the unperturbed electron period

$$\Delta\xi(E_{n+1})\equiv\xi_{n+1}-\xi_n=\pi/\Omega(E_{n+1})\approx\pi/\Omega_{E_{n+1}}, \qquad (27)$$

for the condition $B_0^2\ll\kappa E$. As a result, equations (26, 27) can form a Chirikov-like mapping [23], from which the stochastic condition reads $K_y=|d\xi_{n+1}/d\xi_n-1|=|d\Delta\xi/dE_{n+1}\cdot d\Delta E_y/d\xi_n|\gtrsim 1$ [21, 22], where

$$K_y=k_y\left(E_{max}^{abs}/E\right)^{7/6}\varepsilon>1, \qquad (28)$$

where $k_y=-2^{4/3}\pi^2 Ai'(\eta)$ is a numerical factor. It follows that the decaying of $Ai'(\eta)$ with the increase of $B_0$ will leads a lower boundary for the stochastic electron energy to keep $\eta$ small. The maximum energy is reduced from that without $B_0$ in equation (23) (e.g., see the blue filled region in Fig. 3 where the red curve corresponds to $K_y=1$). If we ignore the numerical factor order of unity, then the stochastic electron motion approximately takes place in the energy region of $(B_0 C_\perp^{-1})^{3/2}<E/4E_{max}^{abs}<\varepsilon^{6/7}$ and, therefore, for $B_0>B_{cri}\equiv C_\perp\varepsilon^{4/7}$ there is no room for stochasticity.

However, for the laser polarized in x-direction, the first term on RHS of equation (25b) is present due to the magnetic field. As a result, this part could slightly increase the energy gain during each kick and thus the maximum stochastic energy shown in equation (23) for $B_0 = 0$. One similar figure with Fig. 3 can be obtained for this case except that the approximately vertical line along the maximum stochastic energy has a tiny curvature toward large E for proper $B_0$. For both cases, the maximum stochastic electron energies are smaller than the low-n resonance energy.

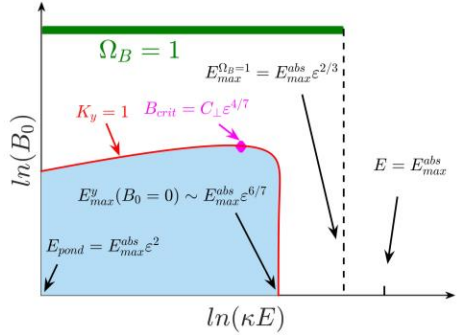

Fig. 3. Schematic view of the electron heating via the overlap of the resonances $n\Omega = 1$ for high-n (stochasticity in the blue filled region) for laser polarized along the static electric field and single low-n resonance of $\Omega_B = 1$ (green bar where the width has no meaning and it covers the energy $0 < E < E_{max}^{\Omega_B=1}$) for both laser polarizations. The numerical factors order of unity for these maximum energies has been omitted.

In order to check these analyses, we performed numerical simulations to solve the 3/2D Hamiltonian equations and display the electron motion in the Poincaré mappings ($E_n, \xi_n$), where the quantities are picked from the center of the nonadiabatic regions as $\tilde{p}_y = 0$. Fig. 4 has shown the results of electron in the laser wave polarized along the transverse electric for $a_0 = 1$, $\kappa = 10^{-5}$, $C_\perp = 1$ and longitudinal magnetic field with (a) $B_0 = 0$, (b) $B_0 = 0.02$, and (c) $B_0 = 0.04$. From Fig. 4(b) we see a lower stochastic energy boundary while the maximum one remains almost unchanged. The termination of the stochasticity occurs around the critical magnetic field of $B_{cri} \approx 0.04$ as shown in Fig. 4(c) (in the simulations, we found that $B_0 = 0.04$ is just below the critical magnetic field and any small increase of $B_0$ will result in regular electron motion). Shown in Fig. 5 are the results for laser polarized along the x-direction with the parameters of $a_0 = 8$, $\kappa = 10^{-4}$, (a) $B_0 = 0$, and (b) $B_0 = 0.05$. We see the increase of the maximum stochastic energy for a proper $B_0$ in Fig. 5(b). For stronger magnetic field, the electron Poincaré mappings are similar to those in Fig. 4 (b, c). However, the difference between the stochastic electron motions due to the laser polarizations could be eliminated by adding relativistic momentum component of $\bar{P}_x \gg 1$.

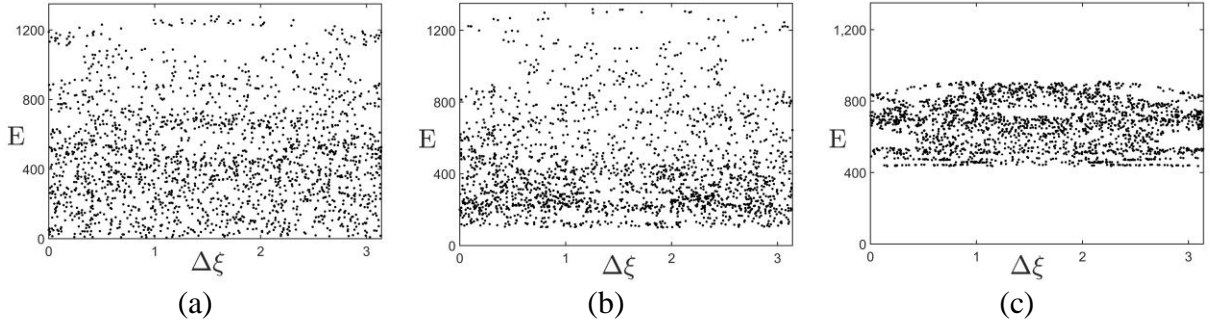

Figure 4. Poincaré mappings for electron moving in the laser wave polarized along the transverse electric for $a_0 = 1$, $\kappa = 10^{-5}$, $C_\perp = 1$ and longitudinal magnetic fields with (a) $B_0 = 0$, (b) $B_0 = 0.02$, and (c) $B_0 = 0.04$, where $\Delta\xi \equiv \xi_n - m\pi$ and $m \equiv [\xi_n/\pi]$ is the largest integer that is smaller than $\xi_n/\pi$.

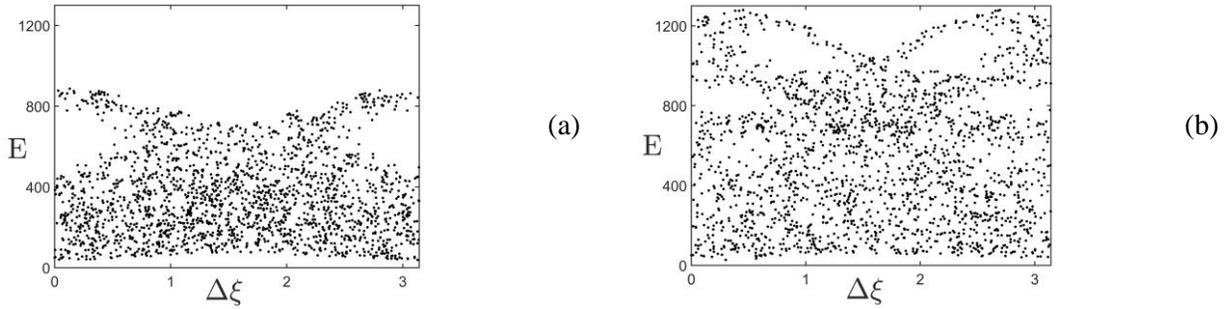

Figure 5. Poincaré mappings for electron moving in the laser wave polarized across to the transverse electric for $a_0 = 8$, $\kappa = 10^{-4}$, $C_\perp = 1$ and longitudinal magnetic fields with (a) $B_0 = 0$ and (b) $B_0 = 0.05$.

## 5. Conclusions

The 3/2 dimensional Hamiltonian method for electrons in the intense laser radiation and quasi-static transverse electric and longitudinal magnetic fields has been derived, within which the electron heating is examined both in the stochastic regime due to the overlap of broadened high harmonic resonances of electron frequency in weak electric and magnetic fields with the laser frequency, and for the low-n resonance for a strong longitudinal magnetic field. For both cases, the maximum electron energies, well beyond the ponderomotive scaling, have been estimated, which depend only on the small parameter ε defined in Eq. (15). In the former case, the presence of a weak longitudinal magnetic field would reduce the maximum stochastic energy for electron in the laser polarized along the static electric field only, whereas it slightly increases the maximum energy for laser across to the electric field when $\bar{P}_x = 0$. It also sets a lower boundary of stochastic electron energy such that the stochasticity will be terminated when the lower boundary meets the upper one (magnetic field exceeds a critical value $B_{crit} \sim C_\perp \varepsilon^{4/7}$). In the latter case, the efficient electron heating is only possible via the first harmonic resonance of $\Omega \approx \Omega_B \approx 1$. The maximum resonant energy, regardless the laser polarization, is above the maximum stochastic energy. Numerical simulations have been performed and the results confirm these analyses.

**Acknowledgments.** This work has been supported by the University of California Office of the President Lab Fee grant number LFR-17-449059.